\documentclass[11pt]{article}
\usepackage[ansinew]{inputenc} 
\usepackage{amssymb,amsfonts}
\usepackage{amsthm}

\begin{document}

\begin{center} 
{\large {\bf Gauge Symmetry Breaking: \\ Higgs-less Mass Generation and 
Radiation Phenomena}}\footnote{Work partially supported by the DGICYT and 
Fundación Séneca under projects BFM 2002-00778 and PB/9/FS/02.} 
\end{center}
\bigskip
\bigskip
\centerline{ {\sc M. Calixto } } 
\bigskip
Departamento de Matemática Aplicada y Estad\'\i stica, Universidad 
Politécnica de Cartagena, Paseo Alfonso XIII 56, 30203 Cartagena, Spain.

\bigskip

\begin{center}
{\bf Abstract}
\end{center}
\small          

\begin{list}{}{\setlength{\leftmargin}{3pc}\setlength{\rightmargin}{3pc}}
\item 
Gauge symmetries generally appear as a constraint algebra, under which one 
expects all physical states to be singlets. However, \emph{quantum 
anomalies} and boundary conditions introduce central charges and change 
this picture, thus causing gauge/diffeomorphism modes to become physical. 
We expose a cohomological ({\it Higgs-less}) generation of mass in 
$U(N)$-gauge invariant Yang-Mills theories through non-trivial representations of the gauge 
group. This situation is also present in black hole evaporation, where the 
Virasoro algebra turns out to be the relevant subalgebra of surface 
deformations of the horizon of an arbitrary black hole. 
\end{list}

\normalsize
\section{Introduction\label{intro}}

Let $T$ be a gauge/diffeomorphism group. From a classical perspective, 
physical states $\Psi_{\rm phys.}$ are expected to be singlets under $T$, 
i.e. \begin{equation}U \Psi_{\rm phys.}=\Psi_{\rm phys.}, 
\;\;U=e^{\varphi^a\Phi_a}\in T.\label{finiteGauss}\end{equation}For 
example, in standard Yang-Mills theory, the infinitesimal counterpart of 
the finite expression (\ref{finiteGauss}) is nothing other than the 
``Gauss law'' condition $\Phi_a\Psi_{\rm phys.}=0$. However, upon 
quantization, anomalies and boundary conditions can change this picture, 
causing gauge/diffeomorphism modes $\varphi^a$ to become physical, so that 
physical states transform non trivially under $T$, \begin{equation}U 
\Psi_{\rm phys.}=D^{(\epsilon)}_T(U)\Psi_{\rm phys.}, 
\end{equation}according to a representation $D^{(\epsilon)}_T$ of $T$ with 
index $\epsilon$. Eventually, the index $\epsilon$ could represent a 
$\vartheta$-angle or a mass parameter $m$ and, in general, it labels {\it 
non-equivalent quantizations}. In fact, the possibility of non-trivial 
representations 
$D^{(m)}_T$ of a $T=U(N)$-invariant Yang-Mills theory will lead to a 
(\emph{Higgs-less}) generation of mass for vector bosons. The mass 
parameters $m$ show up as central charges in the Lie algebra of 
constraints, which transmute to second-class constraints. Some of the 
gauge modes become physical, i.e., they acquire dynamics outside the 
null-mass shell and provide the longitudinal field degrees of freedom that 
massless vector bosons need to form massive vector bosons (see Refs. 
\cite{ymas} and later on Sec. \ref{ymsec}). This seems to be an important 
and general feature of quantum gauge theories as opposite to their 
classical counterparts. In fact, this situation is also present in quantum 
gravity, where the Virasoro algebra turns out to be the relevant 
subalgebra of surface deformations of the horizon of an arbitrary black 
hole and constitutes the general gauge (diffeomorphism) principle that 
governs the density of states. Nevertheless, although surface deformations 
appear as a constraint algebra, under which one might expect all the 
physical states on the horizon to be singlets, \emph{quantum anomalies} 
and boundary conditions introduce central charges and change this picture, 
thus causing gauge/diffeomorphism modes to become physical along the 
horizon (see e.g. \cite{Carlipcqg} and later on Sec. \ref{gwa}). 

In order to set the context, let us describe a simple, but illustrative, 
example of an abstract quantizing algebra $\tilde{{\cal G}}$ which 
eventually applies to a diversity of physical systems. After all, any 
consistent (non-perturbative) quantization is mostly a unitary irreducible 
representation of a suitable (Lie, Poisson) algebra.

\section{A simple abstract quantizing algebra}
Our particular algebra under study will be the following: 
\begin{eqnarray}
\left[X_j,P_k\right]&=&i\delta_{jk}I\,,\nonumber\\ 
\left[\Phi_a,\Phi_b\right]&=&if_{ab}^c\Phi_c+ 
i\epsilon_{ab}I\,,\label{thealgebra}\\ 
\left[X_j,\Phi_a\right]&=&i\check{f}_{ja}^k X_k,\,\,\, 
\left[P_j,\Phi_a\right]=i\check{f}_{ja}^k P_k,\nonumber 
\end{eqnarray} 
where $X_j$ and $P_k$ represent standard ``position'' and ``momentum'' 
operators, respectively, corresponding to the {\it extended phase space} 
${\cal F}\sim\mathbb R^{2m}$ of the preconstrained (free-like) theory; 
The operators $\Phi_a$ represent the constraints which, for the moment, 
are supposed to close a Lie subalgebra 
$\tilde{{\cal T}}$ with structure constants $f_{ab}^c$ and central charges 
$\epsilon_{ab}$. 
We also consider a diagonal action of constraints $\Phi$ on $X$ and $P$ 
with structure constants 
$\check{f}_{ja}^k$  (non-diagonal actions mixing $X$ and $P$ lead to 
interesting ``anomalous'' situations which we shall not discuss here 
\cite{symplin}). By 
$I$ we simply denote the identity operator, that is, the generator of the 
typical phase invariance 
$\Psi\sim e^{i\beta}\Psi$ of Quantum Mechanics. At this stage, it is worth
mentioning that we could have introduced dynamics in our model by adding a 
Hamiltonian operator $H$ to $\tilde{{\cal G}}$. However, we have preferred 
not to include it because, although we could make compatible the dynamics 
$H$ and the constraints $\Phi$, the price could result in an unpleasant 
enlarging of $\tilde{{\cal G}}$, which would make the quantization 
procedure much more involved. Anyway, for us, the {\it true} dynamics 
(that which preserves the constraints) will eventually arise as part of 
the set of {\it good} operators (observables) of the theory 
(\ref{goodeq}).

Note that a flexibility in the class of the constraints has being allowed 
by introducing arbitrary central charges $\epsilon_{ab}$ in 
(\ref{thealgebra}). Thus, the operators $\{\Phi_a\}$ represent a mixed set 
of first- and second-class constraints. Let us denote by ${\cal 
T}^{(1)}=\{\Phi^{(1)}_n\}$ the subalgebra of first-class constraints, that 
is, the ones which do not give rise to central terms proportional to 
$\epsilon_{ab}$ at the right hand side of the commutators (\ref{thealgebra}). The 
rest of constraints (second-class) will be arranged by conjugated pairs 
$(\Phi^{(2)}_{+\alpha},\Phi^{(2)}_{-\alpha})$, so that 
$\epsilon_{+\alpha,-\alpha}\not=0$. 

The simplest (`classical') case is when $\epsilon_{ab}=0,\,\,\forall a,b$, 
that is, when all constraints are first class ${\cal T}^{(1)}={\cal 
T}=\tilde{{\cal T}}/u(1)$ and wave functions are singlets under ${\cal 
T}$. However, the `quantum' case $\epsilon_{ab}\not=0$ entails 
non-equivalent quantizations with important physical consequences. This 
possibility indicates a non-trivial response (\ref{finiteGauss}) of the 
wave function 
$\Psi$ under $\tilde{{\cal T}}$. That is, $\Psi$ 
acquires a non-trivial dependence on extra degrees of freedom 
$\phi^{(2)}_{-\alpha}$ (`negative modes' attached to pairs of 
second-class constraints), in addition to the usual configuration space 
variables $x_j$ (attached to $X_j$). 

Let us formally outline the actual construction of the unitary irreducible 
representations of the group $\tilde{G}$, with Lie-algebra 
(\ref{thealgebra}), following the Group Approach to Quantization framework 
\cite{GAQ}. Wave functions 
$\Psi$ are defined as complex functions on $\tilde{G}$, 
$\Psi:\tilde{G}\rightarrow \mathbb C$, so that the (let us say) left-action 
\begin{equation}L_{\tilde{g}'}\Psi(\tilde{g})\equiv\Psi(\tilde{g}'^{-1} *\tilde{g}),\,\,
\tilde{g}',\tilde{g}\in\tilde{G} \label{repre} 
\end{equation}defines a reducible (in general) representation of $\tilde{G}$. The 
reduction is achieved by means of that maximal set of right restrictions 
on wave functions 
\begin{equation}R_{\tilde{g}_p}\Psi=\Psi,\,\, \forall \tilde{g}_p\in G_p,\,\label{pola} \end{equation}(which 
commute with the left action) compatible with the natural condition 
$I\Psi=\Psi$. The right restrictions (\ref{pola}) generalize the notion of 
{\it polarization conditions} of Geometric Quantization and give rise to a 
certain representation space depending on the choice of the subgroup 
$G_p\subset\tilde{G}$. For the algebra (\ref{thealgebra}), a polarization 
subgroup can be 
$G_p^=F_P\times_sT_p$, that is, the semi-direct product of 
the Abelian group of translations $F_P\sim\mathbb R^{m}$ generated by 
${\cal F}_P\equiv\{P_k\}$ (half of the symplectic generators in ${\cal F}\sim\mathbb R^{2m}$) 
times a polarization subalgebra ${\cal T}_p=\{\Phi^{(1)}_n, 
\Phi^{(2)}_{+\alpha}\}$ of $\tilde{T}$ consisting of first-class 
constraints (the \emph{unbroken gauge subalgebra} ${\cal T}^{(1)}$) and 
half of second-class constraints (namely, the `positive modes' 
$\Phi^{(2)}_{+\alpha}$). The polarization conditions (\ref{pola}) lead to 
the configuration-space representation made of wave functions 
$\Psi(x_j,\phi^{(2)}_{-\alpha})$ depending arbitrarily on the group 
coordinates on $\tilde{G}/G_p$ only. Thus, as mentioned above, wave 
functions transform non-trivially under the left-action 
$L_\phi\Psi(\tilde{g})= D_{\tilde{T}}^{(\epsilon)}(\phi)\Psi(\tilde{g})$ of $\tilde{T}$ according 
to a given representation 
$D_{\tilde{T}}^{(\epsilon)}$ like in (\ref{finiteGauss}). The physical Hilbert space is made 
of those wave functions 
$\Psi_{{{\rm ph.}}}$ that transform as \emph{highest-weight vectors}  
under $\tilde{T}$, that is, they stay invariant under the left-action of 
first-class constraints and (let us say) negative second-class modes: 
\begin{eqnarray} L_{\phi^{(1)}_n}\Psi_{{\rm 
ph.}}\!\!\!\!&=&\!\!\!\! \Psi_{{{\rm ph.}}}\,,\; n=1,..., 
\dim(T^{(1)}),\nonumber\\ L_{\phi^{(2)}_{-\alpha}}\Psi_{{{\rm 
ph.}}}\!\!\!\!&=&\!\!\!\! \Psi_{{{\rm ph.}}} \,, \; 
\alpha=1,...,\dim(T/T^{(1)})/2\,, \label{tpcons} \end{eqnarray} which 
close the subgroup $T_p\subset\tilde{T}$. 

The counting of {\it true degrees of freedom} is as follows: 
polarized-constrained wave functions (\ref{tpcons}) depend arbitrarily on 
$d=\dim(\tilde{G})-\dim(G_p)-\dim(T_p)-1$ reduced-space coordinates (we are 
subtracting the phase coordinate $e^{i\beta}$ too). The algebra of 
observables of the theory, $\tilde{{\cal G}}_{{\rm good}} \subset{\cal 
U}(\tilde{{\cal G}})$ (a subalgebra of the universal enveloping algebra), 
has to be found inside the {\it normalizer} of constraints, that is: 
\begin{equation}\left[\tilde{{\cal G}}_{{\rm good}},{\cal T}_p\right]\subset {\cal T}_p\,.\label{goodeq} 
\end{equation}From this characterization, the subalgebra of first-class constraints 
${\cal T}^{(1)}$ become a horizontal ideal (a {\it gauge} subalgebra 
\cite{config}) of $\tilde{{\cal G}}_{{\rm good}}$. The Hamiltonian 
operator has to be found inside $\tilde{{\cal G}}_{{\rm good}}$ by using 
extra physical arguments. 

\section{Global considerations:  the quantizing group}

In order to discuss some global (versus local) problems in quantization, 
it is necessary to translate the previous infinitesimal (algebraic) 
concepts to their finite counterparts. The exponentiation of the algebra 
(\ref{thealgebra}) leads to a Weyl-symplectic-like group $\tilde{G}$, with 
group law:
\[\begin{array}{rcl} 
({g}'',\zeta'')&=&({g}',\zeta')*({g},\zeta)=(g'g,\zeta'\zeta 
e^{\frac{i}{2\hbar}\xi(g,g')}), \;\; g=(\vec{x},\vec{p},U), \\ 
U''&=&U'U\in {T}\,, \\ \vec{V}''&=&\vec{V}'+U'\vec{V}\in 
F,\;\;\vec{V}=\left(\begin{array}{c}\vec{x} 
\\ \vec{p}\end{array}\right)_{2m\times 1},  \\ \zeta''&=&\zeta'\zeta 
e^{\frac{i}{2\hbar}\xi(g,g')}\in U(1),\;\; \xi(g,g')= \xi(g,g')_{\rm 
B}+\xi(g',g)_\epsilon, \\ \xi(g,g')_{\rm B}&=&\vec{V}'^t 
\left(\begin{array}{cc}0 &I_{m} \\ -I_{m} &0 
\end{array}\right)U'\vec{V},\end{array} 
\]
which is a central extension $\tilde{G}\sim G\times U(1)$ of the 
semidirect product $G=\mathbb R^{2m}\times_s T$ of phase-space 
translations 
$\vec{V}\in \mathbb R^{2m}$ and gauge transformations $U=e^{\phi^a\Phi_a}\in T$ by $U(1)\ni \zeta$. The map  
$\xi:G\times G\rightarrow \mathbb{R}$ is a two-cocycle with two parts: 
1) The Bargmann cocicle $\xi(g,g')_{\rm B}$ says that position 
$\vec{x}$ and momenta $\vec{p}$ are conjugated variables 
[see the first commutator of Eq. (\ref{thealgebra})], and 2) the 
two-cocicle 
$\xi(g',g)_\epsilon$ is meant to provide couples of second-class 
constraints. 

Two two-cocycles are said to be equivalent if they differ by a coboundary, 
i.e. a two-cocycle which can be written in the form 
$\xi(g',g)=\eta(g'*g)- \eta(g')-\eta(g)$, where $\eta(g)$ is called the 
generating function of the coboundary. Although two-cocycles differing by 
a coboundary lead to equivalent central extensions as such, there are some 
coboundaries which provide a non-trivial connection \begin{equation}
\Theta=\left.\frac{\partial}{\partial g^j}\xi(g',g)\right|_{g'=g^{-1}}dg^j 
-i\hbar\zeta^{-1}d\zeta\,,\label{thetagen} \end{equation}on the fibre 
bundle $\tilde{G}$, and Lie-algebra structure constants different from 
those of the direct product $G\times U(1)$. These are generated by a 
function 
$\eta$ with a non-trivial gradient at the 
identity $\left.d\eta(g)\right|_{g=e}= 
\left.\frac{\partial\eta(g)}{\partial g^j}\right|_{g=e}d g^j\not=0$, and 
can be divided into pseudo-cohomology equivalence subclasses (see 
\cite{julio} in this volume). Pseudo-cohomology plays an important role in 
the theory of finite-dimensional semi-simple groups, as they have trivial 
cohomology. For them, pseudo-cohomology classes are associated with 
coadjoint orbits (see \cite{julio}). Next section, we shall show how the 
introduction of coboundaries in some physical systems alters the 
corresponding quantum theory. From the mathematical point of view, 
pseudo-cocicles entail trivial redefinitions of some Lie-algebra 
generators; however, from the physical point of view, they resemble the 
appearance of \emph{non-zero vacuum expectation values}: 
\begin{equation}
\langle 0|(\Phi_a-\epsilon_a {I})|0\rangle=0\Rightarrow \langle 
0|\Phi_a|0\rangle =\epsilon_a. \label{nzvev}
\end{equation}
Let us discuss inside this framework the quantization of massless and 
massive electromagnetism, linear Gravity, Abelian two-form and non-Abelian 
Yang-Mills gauge field theories, and to point out a cohomological 
(Higgs-less) origin of mass.

\section{Unified quantization of massless and \break massive vector and 
tensor bosons} 
\subsection{The electromagnetic and Proca fields:}
Let us start with the simplest case of the electromagnetic field. Let us 
use a Fourier parametrization \[A_{\mu}(x)\equiv \int 
\frac{d^3k}{2k^0}[a_{\mu}(k)e^{-ikx}+ a^\dag_{\mu}(k)e^{ikx}]\,,\;\; 
\Phi(x)\equiv \int \frac{d^3k}{2k^0}[ \varphi(k)e^{-ikx}+ 
 \varphi^\dag(k)e^{ikx}]\,,\] for the vector potential $A_\mu(x)$ and the constraints $\Phi(x)$ (the 
generators of local $U(1)(x)$ gauge transformations). The Lie algebra 
$\tilde{{\cal G}}$ of the quantizing electromagnetic group $\tilde{G}$ has the 
following form \cite{ymas} \begin{eqnarray*}
\left[a_{\mu}(k),a^\dag_{\nu}(k')\right]= 
 \eta_{\mu\nu}\Delta_{kk'}I\,, & & 
 \left[\varphi(k),
\varphi^\dag(k')\right]=k^2\Delta_{kk'}I\,, \\ 
\left[a^\dag_{\mu}(k),\varphi(k')\right]= -ik_{\mu}\Delta_{kk'}I \,,& & 
\left[a_{\mu}(k), \varphi^\dag(k')\right]=-ik_{\mu}\Delta_{kk'}I \,,  
\end{eqnarray*} where $\Delta_{kk'}= 2k^0 \delta^3(k-k')$ is the generalized delta 
function on the positive sheet of the mass hyperboloid and $k^2=m^2$ is 
the squared mass. Constraints are first-class for $k^2=0$ and constraint 
equations 
$\varphi\Psi=0=\varphi^\dag\Psi$ keep 2 field degrees of freedom out of 
the original 4, as corresponds to a photon. For $k^2\not=0$, constraints 
are second-class and the restrictions $\varphi\Psi=0$ keep 3 field degrees 
of freedom out of the original 4, as corresponds to a Proca field. 
\subsection{Linear gravity:}
For symmetric and anti-symmetric tensor potentials $A^{(\pm)}_{\mu\nu}$, 
the algebra is the following \cite{ymas}: 
\begin{eqnarray*}\left[a^{(\pm)}_{\lambda\nu}(k), a^{\dag(\pm)}_{\rho\sigma}(k')\right]= 
 N_{\lambda\nu\rho\sigma}^{(\pm)}\Delta_{kk'}I\,, \;\;
\left[\varphi^{(\pm)}_{\rho}(k), 
\varphi^{\dag(\pm)}_{\sigma}(k')\right]=k^2M^{(\pm)}_{\rho\sigma}(k) 
\Delta_{kk'}I\,,\\ 
\left[a^{\dag(\pm)}_{\lambda\nu}(k),\varphi^{(\pm)}_{\sigma}(k')\right]= 
-ik^{\rho}N_{\lambda\nu\rho\sigma}^{(\pm)}\Delta_{kk'}I\,,\;\; 
 \left[a^{(\pm)}_{\lambda\nu}(k),
\varphi^{\dag(\pm)}_{\sigma}(k')\right]=-ik^{\rho}N_{\lambda\nu\rho\sigma}^{(\pm)}\Delta_{kk'}I\,, 
 \end{eqnarray*} where $M^{(\pm)}_{\rho\sigma}(k) \equiv \eta_{\rho\sigma} 
- \kappa_{(\mp)}\frac{k_\rho k_\sigma}{k^2}$ and 
$N_{\lambda\nu\rho\sigma}^{(\pm)} \equiv \eta_{\lambda\rho}\eta_{\nu\sigma} \pm   
\eta_{\lambda\sigma}\eta_{\nu\rho} - 
\kappa_{(\pm)}\eta_{\lambda\nu}\eta_{\rho\sigma}$, with $\kappa_{(+)}=1$ 
and 
$\kappa_{(-)}=0$. For the massless 
$k^2=0$ case, all constraints are first-class for the symmetric case, 
whereas the massless, anti-symmetric case possesses a couple of 
second-class constraints: 
\begin{equation}\left[\check{k}^\rho\varphi^{(-)}_{\rho}(k), 
{\check{k}'}{}^\sigma\varphi^{(-)\dag}_{\sigma}(k')\right]=4(k^0)^4\Delta_{kk'} 
I\,,\label{2ndclass} \end{equation}where $\check{k}^\rho\equiv k_\rho$. 
Thus, first-class constraints for the massless anti-symmetric case are 
${\cal T}_{(-)}^{(1)}= \{\epsilon_\mu^\rho\varphi^{(-)}_{\rho},\, 
\epsilon_\mu^\rho\varphi^{(-)\dag}_{\rho}\},\, \mu=0,1,2,\,$ where 
$\epsilon_\mu^\rho$ is a tetrad which diagonalizes the matrix 
$P_{\rho\sigma}=k_\rho k_\sigma$; in particular, we choose 
$\epsilon_3^\rho\equiv \check{k}^\rho$ and 
$\epsilon_0^\rho\equiv k^\rho$. There are $2=10-8$ true degrees 
of freedom for the symmetric case (a massless graviton) and $1=6-5$ for 
the anti-symmetric case (a pseudo-scalar particle). 

For $k^2\not=0$, all constraints are second-class for the symmetric case, 
whereas, for the anti-symmetric case, constraints close a Proca-like 
subalgebra which leads to three pairs of second-class constraints, and a 
pair of gauge vector fields $(k^\lambda\varphi^{(-)}_{\lambda},\, 
k^\lambda\varphi^{(-)\dag}_{\lambda})$. The constraint equations keep 
$6=10-4$ field degrees of freedom for the symmetric case 
(massive spin 2 particle + massive scalar field ---the trace of the 
symmetric tensor), and $3=6-3$ field degrees of freedom for the 
anti-symmetric case (massive pseudo-vector particle). 

\subsection{$SU(N)$-Gauge Invariant Yang-Mills Theories:\label{ymsec}} 
Let us show how mass can enter Yang-Mills theories through central 
(pseudo) extensions of the corresponding gauge group. This mechanism does 
not involve extra (Higgs) scalar particles and could provide new clues for 
the better understanding of the nature of the Symmetry Breaking Mechanism. 
We are going to outline the essential points and refer the interested 
reader to the Ref. \cite{ymas} for further information. 

Let us denote by $A^\mu(x)=r^a_bA^\mu_a(x)T^b,\,\mu=0,...,3; 
a,b=1,...,N^2-1={\rm dim}(SU(N))$ the Lie-algebra valued vector potential 
attached to a non-Abelian gauge group which, for simplicity, we suppose to 
be unitary, say $T={\rm Map}(\mathbb R^4,SU(N))= 
\{U(x)=\exp{\varphi_a(x)T^a}\}$, where $T_a$ are the generators of 
$SU(N)$, which satisfy the commutation relations 
$[T_a,T_b]=C_{ab}^cT_c$, and the coupling constant matrix $r^a_b$ reduces 
to a multiple of the identity 
$r^a_b=r\delta^a_b$. 
We shall also make partial use of the gauge freedom to set the temporal 
component $A^0=0$, so that the Lie-algebra valued electric field is simply 
$E^j(x)\equiv r^a_bE^j_a(x)T^b=-\dot{A}^j(x)$. In this case, there is 
still a residual gauge invariance $T={\rm Map}(\mathbb R^3,SU(N))$. 

The proposed (infinite dimensional) quantizing group for quantum 
Yang-Mills theories will be a central extension $\tilde{G}$ of 
$G=(G_A\times G_E)\times_s T$ (semi-direct product of the cotangent 
group of the Abelian group of Lie-algebra valued vector potentials and the 
non-Abelian gauge group $T$) by $U(1)$. More precisely, the group law for 
$\tilde{G}$, $\tilde{g}''=\tilde{g}*\tilde{g}$, with $\tilde{g}=(A^j_a(x),E^j_a(y),U(z);\zeta)$, 
 can be explicitly written as (in natural 
units $\hbar=1=c$): \begin{eqnarray}U''(x)&=&U'(x)U(x)\,,\nonumber\\ 
\vec{A}''(x)&=&\vec{A}'(x)+U'(x)\vec{A}(x)U'(x)^{-1}\,,\nonumber\\ 
\vec{E}''(x)&=&\vec{E}'(x)+U'(x)\vec{E}(x)U'(x)^{-1}\,,\nonumber\\ 
\zeta''&=&\zeta'\zeta\exp\left\{-\frac{i}{r^2}\sum_{j=1}^2 
\xi_j(\vec{A}',\vec{E}',U'|\vec{A},\vec{E},U)\right\}\,; \label{ley}\\ 
\xi_1(g'|g)&\equiv& \int{{d}^3x\,{\rm tr}\left[\,\left(\begin{array}{cc} 
\vec{A}' & \vec{E}'\end{array}\right) W \left(\begin{array}{c} 
U'\vec{A}U'^{-1} \\ U'\vec{E}U'^{-1} 
\end{array}\right)\right]}\,,\nonumber\\ \xi_2(g'|g)&\equiv& 
\int{{d}^3x\,{\rm tr}\left[\,\left(\begin{array}{cc} \nabla U'U'^{-1} & 
\vec{E}'\end{array}\right) W \left(\begin{array}{c} U'\nabla 
UU^{-1}U'^{-1} \\ U'\vec{E}U'^{-1} 
\end{array}\right)\,\right]}\,,\nonumber \end{eqnarray} \noindent where 
$W=\left(\begin{array}{cc} 0 & 1 \\ -1 & 0\end{array}\right)$ is a symplectic matrix and we 
have split up the cocycle $\xi$ into two distinguishable and typical 
cocycles $\xi_j,\,\,j=1,2$. The first cocycle 
$\xi_1$ (a Bargmann-like one) is meant to provide {\it dynamics} 
for the vector potential, so that the couple $(A,E)$ corresponds to a 
canonically-conjugate pair of field coordinates. The second cocycle 
$\xi_2$, the {\it mixed} cocycle, provides a non-trivial (non-diagonal) 
action of the gauge subgroup $\tilde{T}$ on vector potentials and 
determines the number of degrees of freedom of the constrained theory; in 
fact, it represents the ``quantum'' counterpart of the ``classical'' 
unhomogeneous term 
$U(x)\nabla U(x)^{-1}$ we miss at the right-hand side of the gauge 
transformation of $\vec{A}$ (second line of (\ref{ley})), that is, the 
vector potential $\vec{A}$ has to transform homogeneously under the action 
of the gauge group $T$ in order to define a proper group law, whereas the 
inhomogeneous term $U(x)\nabla U(x)^{-1}$ modifies the {\it phase} $\zeta$ 
of the wave function according to $\xi_2$ (see \cite{ymas} for a covariant 
form of this ``quantum'' transformation). 

To make more explicit the intrinsic significance of these two quantities 
$\xi_j\,,\,\, j=1,2$, let us compute the non-trivial Lie-algebra 
commutators of the right-invariant vector fields (that is, the generators 
of the left-action 
$L_{\tilde{g}'}(\tilde{g})=\tilde{g}'*\tilde{g}$ of $\tilde{G}$ on itself) from  the group law (\ref{ley}). 
They are explicitly: \begin{eqnarray}\left[\hat{A}^j_a(x), 
\hat{E}^k_b(y)\right]&=& 
i\delta_{ab}\delta^{jk}\delta(x-y)\hat{I}\,,\nonumber\\ 
\left[\hat{A}^j_a(x), \hat{\varphi}_b(y)\right]&=&-iC_{ab}^c\delta(x-y) 
\hat{A}^j_c(x) 
-\frac{i}{r}\delta_{ab}\partial^j_x\delta(x-y)\hat{I}\,,\label{YM}\\ 
\left[\hat{E}^j_a(x), \hat{\varphi}_b(y)\right]&=&-iC_{ab}^c\delta(x-y) 
\hat{E}^j_c(x)\nonumber\\ \left[\hat{\varphi}_a(x), 
\hat{\varphi}_b(y)\right]&=&-iC_{ab}^c\delta(x-y) \hat{\varphi}_c(x) 
\,.\nonumber 
\end{eqnarray} 

The unitary irreducible representations of $\tilde{G}$ with structure 
subgroup 
$\tilde{T}=T\times U(1)$ (a direct product for this case) represent a quantum 
theory of 
$n=N^2-1={\rm dim}(SU(N))$ 
interacting massless vector bosons. Indeed, we start with $f=3n$ field 
degrees of freedom, corresponding to the basic operators 
$\{\hat{A}^j_a(x),\hat{E}^j_a(x)\}$ 
(the ones that have a conjugated counterpart); the constraints 
(\ref{tpcons}) provide $c=n$ independent restrictions 
$\hat{\varphi}_a(x)\psi=0,\,\, a=1,\dots, n$ (the quantum implementation 
of the non-Abelian Gauss law), since they are first-class constraints and 
we choose the trivial representation 
$D_{\tilde{T}}^{(\epsilon)}(\tilde{g}_t)=1,\,\,\forall \tilde{g}_t=(0,0,U(x);1)\in T$, 
restrictions which lead to $f_c=f-c=2n$ field degrees of freedom 
corresponding to an interacting theory of $n$ massless vector bosons. 

However, more general representations 
$D_{\tilde{T}}^{(\epsilon)}(U)=e^{i\epsilon_{{}_U}}$ 
can be considered when we impose additional boundary conditions like $U(x) 
\stackrel{x\rightarrow\infty}{\longrightarrow}\pm I$, that is, when we 
compactify the space $\mathbb R^3\rightarrow S^3$ so that the gauge group 
$T$ falls into disjoint homotopy classes $\{U_l\,,\, 
\epsilon_{{}_{U_l}}\equiv l\vartheta\}$ labeled by integers $l\in \mathbb 
Z=\pi_3(SU(N))$ (the third homotopy group). The index $\vartheta$ (the 
{\it 
$\vartheta$-angle}) parametrizes {\it non-equivalent quantizations}, as 
the Bloch momentum $\epsilon$ does for particles in periodic potentials, 
where the wave function acquires a phase 
$\psi(q+2\pi)=e^{i\epsilon}\psi(q)$ after a translation of, let us say, 
$2\pi$. The phenomenon of non-equivalent quantizations can also be 
reproduced by keeping the constraint condition 
$D_{\tilde{T}}^{(\epsilon)}(U)=1$ unchanged at the price of introducing a 
new (pseudo) cocycle 
$\xi_\vartheta$ which is added to the 
previous cocycle $\xi=\xi_1+\xi_2$ in (\ref{ley}). The generating function 
$\eta_\vartheta$ of $\xi_\vartheta$ is 
\begin{equation}\eta_\vartheta(g)=\vartheta\int{d^3x\, {\cal C}^0(x)}\,,\;\;\;\; 
 {\cal C}^\mu=-\frac{1}{16\pi^2}\epsilon^{\mu\alpha\beta\gamma}{\rm tr}
({\cal F}_{\alpha\beta}{\cal A}_\gamma-\frac{2}{3}{\cal A}_\alpha {\cal 
A}_\beta {\cal A}_\gamma)\,, \end{equation}where ${\cal A}\equiv A+\nabla 
UU^{-1}$ and ${\cal C}^0$ is the temporal component of the {\it 
Chern-Simons secondary characteristic class} ${\cal C}^\mu$, which is the 
vector whose divergence equals the Pontryagin density ${\cal P} = 
\partial_\mu {\cal C}^\mu = -\frac{1}{16\pi^2} {\rm tr} ({}^*{\cal 
F}^{\mu\nu} {\cal F}_{\mu\nu})$. Like some total derivatives (namely, the 
Pontryagin density), which do not modify the classical equations of motion 
when added to the Lagrangian but have a non-trivial effect in the quantum 
theory, pseudo-cocycles like 
$\xi_\vartheta$ give rise to non-equivalent quantizations when 
the topology of the space is affected by the imposition of certain 
boundary conditions (``compactification of the space''), even though they 
are trivial cocycles of the ``unconstrained'' theory. The phenomenon of 
non-equivalent quantizations can be also sometimes understood as a {\it 
Aharonov-Bohm-like effect} (an effect experienced by the quantum particle 
but not by the classical particle) and 
$d\eta(g)=\frac{\partial\eta(g)}{\partial g^j}d g^j$ can be seen 
as an {\it induced gauge connection} (see \cite{FracHall} for the example 
of a super-conducting ring threaded by a magnetic flux) which modifies 
momenta according to the minimal coupling. 

We can also go further and consider more general representations 
$D_{\tilde{T}}^{(\epsilon)}$ of $\tilde{T}$ (in particular, non-Abelian representations) 
by adding extra pseudo-cocycles to $\xi$. This is the case of 
\begin{equation}\xi_\lambda(g'|g)\equiv -2\int{{d}^3x\, {\rm tr}[ \lambda\left(\log 
(U'U)-\log U'-\log U\right)]}\,,\nonumber \end{equation}which is generated 
by 
$\eta_\lambda(g)=-2\int{{d}^3x\, {\rm tr}[ \lambda\log U]}$, where 
$\lambda=\lambda^aT_a$ is a (mass) matrix carrying some parameters $\lambda^a$ 
which actually characterize the representation of 
$\tilde{G}$. In fact, this pseudo-cocycle alters the gauge group commutators 
and leads to the appearance of new central terms at the right-hand side of 
the last equation in (\ref{YM}), more explicitly: 
\begin{equation}\left[\hat{\varphi}_a(x), \hat{\varphi}_b(y)\right]=-iC_{ab}^c\delta(x-y) 
\hat{\varphi}_c(x) 
-iC_{ab}^c\frac{\lambda_c}{r^2}\delta(x-y)\hat{I}\,.\label{masin} 
\end{equation}
Let us denote by 
$c\equiv{\rm dim}(T^{(1)})$ and $\tau\equiv N^2-1$ the dimensions of the 
rigid subgroup of first-class constraints and $SU(N)$, respectively. 
Unpolarized wave functions 
$\Psi(A^j_a,E^j_a,\phi_a)$ depend on $n=2\times 3\tau+\tau$ field 
coordinates in $d=3$ dimensions; polarization equations (\ref{pola}) 
introduce $p=c+ \frac{n-c}{2}$ independent restrictions on wave functions, 
corresponding to $c$ non-dynamical coordinates in $T^{(1)}$ and half of 
the dynamical ones; finally, constraints (\ref{tpcons}) impose 
$q=c+\frac{\tau-c}{2}$ additional restrictions which leave 
$f=n-p-q=2c+3(\tau-c)$ field degrees of freedom (in $d=3$). 
 These fields correspond to 
$c$ massless vector bosons (2 polarizations) attached to ${T}^{(1)}$ and 
$\tau-c$ massive vector bosons.  In particular, for the massless 
case, we have $c=\tau$, since constraints are {\it first-class} (that is, 
we can impose $q=\tau$ restrictions) and constrained wave functions have 
support on $f_{m=0}=3\tau-\tau=2\tau\leq f_{m\not=0}$ arbitrary fields 
corresponding to $\tau$ massless vector bosons. The subalgebra ${\cal 
T}^{(1)}$ corresponds to the unbroken gauge symmetry of the constrained 
theory. There are distinct symmetry-breaking patterns $\tilde{\cal 
T}\rightarrow {\cal T}^{(1)}$ according to the different choices of 
mass-matrices 
$\lambda_{ab}=f^c_{ab}\lambda_c$ in (\ref{YM}).
 
As already stated in (\ref{nzvev}), pseudo-cocycle parameters such as 
$\lambda_c$ are usually hidden in a redefinition of the generators 
involved in the pseudo-extension 
$\hat{\varphi}_c(x)+\frac{\lambda_c}{r^2}\equiv \hat{\varphi}_c'(x)$, as it 
happens for example with the parameter $c'$ in the Virasoro algebra 
(\ref{vira}), which is a redefinition of ${L}_0$. However, whereas the 
vacuum expectation value $\langle 0_\lambda| 
\hat{\varphi}_c(x)|0_\lambda\rangle$ is zero, the vacuum expectation value 
$\langle 0_\lambda| \hat{\varphi}_c'(x)|0_\lambda\rangle=\lambda_c/r^2$ of 
the redefined operators $\hat{\varphi}_c'(x)$ is non-zero and proportional 
to the cubed mass $\lambda_c\sim m^3_c$ in the `direction' $c$ of the 
unbroken gauge symmetry 
$T^{(1)}$, which depends on the particular choice of the mass matrix 
$\lambda$. Thus, the effect of the pseudo-extension manifests also in a 
different choice of a vacuum in which some gauge operators have a non-zero 
expectation value. This fact reminds us of the Higgs mechanism in 
non-Abelian gauge theories, where the Higgs fields point to the direction 
of the non-zero vacuum expectation values. However, the spirit of the 
Higgs mechanism, as an approach to supply mass, and the one discussed here 
are different, even though they share some common features. In fact, we 
are not making use of extra scalar fields in the theory to provide mass to 
the vector bosons, but it is the gauge group itself that acquires dynamics 
for the massive case and transfers field degrees of freedom to the vector 
potentials to form massive vector bosons. Thus, the appearance of mass 
seems to have a {\it cohomological origin}, beyond any introduction of 
extra scalar particles (Higgs bosons). The full physical implications of 
this alternative approach deserve further study, although some important 
steps have been already done (see \cite{ymas}). 

Also, it would be worth exploring the richness of the case 
$SU(\infty)$ (infinite number of colours), the Lie-algebra of which 
is related to the (infinite-dimensional) Lie-algebra of area preserving 
diffeomorphisms of the torus $SDiff(T^2)$: 
\begin{equation}\left[L_{\vec{m}},L_{\vec{n}}\right]=(\vec{m}\times\vec{n})L_{\vec{m}+\vec{n}}+ 
\vec{\lambda}\cdot\vec{m}\delta_{\vec{m}+\vec{n},0}\hat{I},\;\;\;\; 
\vec{m},\vec{n}\in \mathbb Z\times \mathbb Z \end{equation} which here 
appears centrally extended ($\vec{\lambda}=(\lambda_1,\lambda_2)$ stands 
for the central extension parameter)
\section{String theory and radiation phenomena \label{gwa}} We find also in String 
Theory that the appearance of central terms in the constraint (Virasoro) 
subalgebra 
\begin{equation}
\left[{L}_n,{L}_m\right]=\hbar(n-m){L}_{n+m}+ 
\frac{\hbar^2}{12}(cn^3-c'n)\delta_{n,-m}{I}\,.\label{vira} 
\end{equation} 
does not spoil gauge invariance but forces us to impose a polarization 
subgroup $T_p$ of $\tilde{T}$ only (namely, the `positive modes' 
${L}_{n\geq 0}$) as restrictions $L_n\Psi_{phys}=0$ on physical wave 
functions; for this case, constraints are said to be of second-class. 

Moreover, this situation shows up too in black hole thermodynamics. A 
statistical mechanical explanation of black hole thermodynamics in terms 
of counting of microscopic states has been recently given in 
\cite{Carlipcqg}. According to this reference, there is strong evidence 
that conformal field theories provide a universal (independent of details 
of the particular quantum gravity model) description of low-energy black 
hole entropy, which is only fixed by symmetry arguments. The Virasoro 
algebra turns out to be the relevant subalgebra of surface deformations of 
the horizon of an arbitrary black hole and constitutes the general gauge 
(diffeomorphism) principle that governs the density of states. As already 
said, although surface deformations appear as a constraint algebra, under 
which one might expect all the physical states on the horizon to be 
singlets, quantum anomalies and boundary conditions introduce central 
charges and change this picture, thus causing gauge/diffeomorphism modes 
to become physical along the horizon ---this situation resembles the above 
Higgs-less mechanism of mass generation in Yang-Mills theories. In this 
way, the calculation of thermodynamical quantities, linked to the 
statistical mechanical problem of counting microscopic states, is reduced 
to the study of the representation theory and central charges of a 
relevant symmetry algebra. 

Unruh effect (vacuum radiation in uniformly accelerated frames) is another 
interesting physical phenomenon linked to the previous one. A statistical 
mechanical description (from first principles) of it has also been given 
in \cite{conforme} and related to the dynamical breakdown of part of the 
conformal symmetry $SO(4,2)$: the special conformal transformations 
(usually interpreted as transitions to a uniformly relativistic 
accelerated frame), in the context of a 
$SO(4,2)$ conformally invariant quantum field theory. Unruh effect can be 
considered as a ``first order effect'' that gravity has on quantum field 
theory, in the sense that transitions to uniformly accelerated frames are 
just enough to account for it. To account for higher-order effects one 
should consider more general diffeomorphism algebras. In Refs. 
\cite{infdimal}, the author has introduced higher-$U(N_+,N_-)$-spin 
extensions and higher-dimensional analogies of the infinite 
two-dimensional conformal symmetry (\ref{vira}), generalizing the standard 
$W_\infty$ algebra (a higher-conformal-spin extension of the Virasoro algebra), 
viewed as a tensor operator algebra of $SU(1,1)$ in a 
group-theoretic framework. These centrally-extended infinite-dimensional 
Lie algebras could be useful as potential gauge guiding principles towards 
the formulation of gravity models in realistic dimensions.

\end{document}